\documentclass[conference]{IEEEtran}
\IEEEoverridecommandlockouts
\usepackage{cite}
\usepackage{amsmath,amssymb}

\DeclareMathOperator*{\argmin}{arg\,min}
\usepackage{algorithmic}
\usepackage{graphicx}
\usepackage{caption}
\usepackage{subcaption}
\usepackage{textcomp}
\usepackage{xcolor}
\usepackage{nicefrac}
\usepackage{bm}
\usepackage{hyperref}
\usepackage{tabularx}
\usepackage{booktabs}

\def\BibTeX{{\rm B\kern-.05em{\sc i\kern-.025em b}\kern-.08em
    T\kern-.1667em\lower.7ex\hbox{E}\kern-.125emX}}

\begin{document}

\onecolumn 
{\Huge \bfseries  IEEE Copyright Notice}

\vspace{20mm}

\large This work has been submitted to the IEEE for possible publication. Copyright may be transferred without notice, after which this version may no longer be accessible.

\newpage

\twocolumn

\title{Diagnosis Of Takotsubo Syndrome By Robust Feature Selection From The Complex Latent Space Of DL-based Segmentation Network}

\author{Fahim Ahmed Zaman, Wahidul Alam, Tarun Kanti Roy, Amanda Chang, Kan Liu and Xiaodong Wu
\thanks{Fahim Ahmed Zaman and Xiaodong Wu are with the department of Electrical and Computer engineering, University of Iowa, Iowa City, IA 52242, USA. (email: {fahim-zaman, xiaodong-wu}@uiowa.edu)}
\thanks{Wahidul Alam is with the department of Biomedical engineering, University of Iowa, Iowa City, IA 52242, USA. (email: mohammadwahidul-alam@uiowa.edu)}
\thanks{Tarun Kanti Roy is with the department of Computer Science, University of Iowa, Iowa City, IA 52242, USA. (email: tarunkanti-roy@uiowa.edu)}
\thanks{Amanda Chang is with the Division of Cardiology, Department of Internal Medicine, University of Iowa, Iowa City, IA 52242, USA (e-mails: amanda-chang@uiowa.edu)}
\thanks{Kan Liu is with the Division of Cardiology and Heart and Vascular Center, School of Medicine, Washington University in St Louis, St. Louis, MO 63130, USA (e-mails: kanl@wustl.edu)}
}

\maketitle

\begin{abstract}
Researchers have shown significant correlations among segmented objects in various medical imaging modalities and disease related pathologies. Several studies showed that using hand crafted features for disease prediction neglects the immense possibility to use latent features from deep learning (DL) models which may reduce the overall accuracy of differential diagnosis. However, directly using classification or segmentation models on medical to learn latent features opt out robust feature selection and may lead to overfitting. To fill this gap, we propose a novel feature selection technique using the latent space of a segmentation model that can aid diagnosis. We evaluated our method in differentiating a rare cardiac disease: Takotsubo Syndrome (TTS) from the ST elevation myocardial infarction (STEMI) using echocardiogram videos (echo). TTS can mimic clinical features of STEMI in echo and extremely hard to distinguish. Our approach shows promising results in differential diagnosis of TTS with 82\% diagnosis accuracy beating the previous state-of-the-art (SOTA) approach. Moreover, the robust feature selection technique using LASSO algorithm shows great potential in reducing the redundant features and creates a robust pipeline for short- and long-term disease prognoses in the downstream analysis.
\end{abstract}

\begin{IEEEkeywords}
Segmentation latent space, Feature selection, Takotsubo Syndrome, Echocardiogram video
\end{IEEEkeywords}

\section{Introduction}
Quick diagnosis and accurate treatment by giving proper medication to patients is necessary for life threatening diseases such as acute myocardial infarction (AMI). 
But TTS can mimic clinical and electrocardiographic (ECG) features of AMI and hard to distinguish between them using just echo. Use of angiogram is the gold standard of distinguishing between these two diseases, which is not only invasive, but also slow in process that may endanger the patients in the emergency room. 
Recently deep learning models are studied for the classification of TTS and ST elevation myocardial infarction (STEMI) using echo \cite{zaman,jama}. 
In our earlier study, we demonstrated that using deep learning model as a binary classifier between the two diseases can significantly improve the detection accuracy compared to the physicians and help them make the judgement calls in difficult cases \cite{zaman}. 
Despite having good classification accuracy, DCNN classifiers do not guarantee robust feature selection, particularly in a noisy dataset such as echo. Moreover, artifacts and speckle noise in the echos can generate irrelevant and wrong features that may reduce the overall accuracy.

Looking at the feature maps of the trained classifier, we identified that the basal septal, antero-lateral walls and the apex of the heart are extremely important in decision making of the deep learning models, but not consistent in all the individual cases as they can be affected by echo artifacts. We further investigated and found out that there are indeed significant motion differences in the above mentioned regions of the heart between the two diseases \cite{zaman2023deep}. The pathological evidences also suggest that the Left Ventricle (LV) overall plays a vital role in differentiating these two diseases \cite{assad2022takotsubo, ahmad2017takotsubo,akashi2008takotsubo,yalta2018left,lyon2016current}. But it is impossible to distinguish the spatial and temporal features of the classifier due to the black box nature of the model. Intuitively, the segmentation related features are more robust and correlate the diseases of interest. This inspired us to search for more robust features related to the geometric shape and topology of the LV.

\begin{figure*}[h]
	\centering
	\includegraphics[width=5.5in]{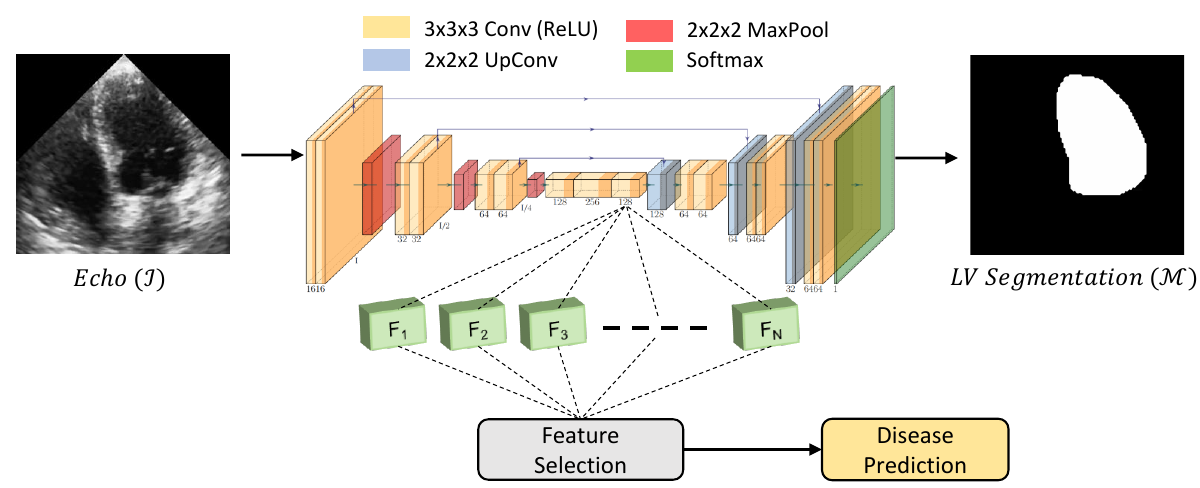}
	\caption{Method workflow diagram. A U-Net shaped architecture is used for LV segmentation from echo. Note that, the echo dataset is a video dataset (2D+t). $\mathcal{I}$ is the input video and $\mathcal{M}$ is the corresponding segmentation mask. Here, only a frame is shown for visual convenience. Latent features from the final convolutional layer of the encoder are used for feature selection. Feature selection block is detailed in \ref{Ft_reduction}. The selected features from feature selection step are used to train binary classifier for disease prediction.}
	\label{Fig2}
\end{figure*}

Baek et al., showed that CNNs trained to perform tumor segmentation task, with no other information than the physician contours, identify a rich set of survival-related image features with remarkable prognostication in a retrospective study \cite{baek}. The authors were able to identify strong correlation between segmentation related features and the disease outcomes with no dependency on any other clinical information. Same trend has also been observed from the validation over an external dataset. Their proposed survival prediction framework composed of two major units: (1) A segmentation network and (2) A survival prediction model. The learned segmentation related features of the latent space was clustered in an unsupervised manner. Then, a logistic regression model is trained for survival prediction. 

Motivated from the survival prediction model, we propose a robust feature selection technique leveraging on the learned latent features of the LV segmentation model that improves diagnosis accuracy. The contribution of this work can be summarized, as follows.

\begin{enumerate}
    \item We propose to use a segmentation model to learn geometric and motion related features of the LV, which allows robust feature selection for differential diagnosis of the TTS based on the pathological findings.
    \item We propose a feature reduction strategy for selecting robust features to improve diagnosis accuracy and reduce computational complexity. Our proposed model shows promising results with the new SOTA diagnosis accuracy for TTS diagnosis.
\end{enumerate}

\section{Dataset}
We used an echocardiogram video (echo) dataset from The University of Iowa hospitals. The echos are acquired by Transthoracic echocardiography (TTE) using standard techniques of 2D echocardiography following the guidelines of the American Society of Echocardiography \cite{lang2015recommendations}. All the echos have the standard apical 4-chamber LV focused view. In total, the selected dataset contains 300 echos (140 of TTS patients, 160 of STEMI patients), having $18 \sim 112$ frames per video. The LVs were manually traced fully by an expert using ITK-snap. The data has been pre-processed by adapting ROI selection technique of \cite{zaman}. 



\section{Methodology}

We propose a deep learning framework that first learns the segmentation related features from LV, then selects only the robust features for differential diagnosis of TTS. The method workflow can be divided into 4 major steps: (1) LV segmentation: A U-net \cite{unet} shaped auto-encoder is trained for LV segmentation, (2) Feature extraction: The features are extracted from the output of the encoder of the segmentation model, (3) Feature selection: Robust features are selected from the extracted latent features of the auto-encoder, (4) Disease prediction: Binary classifier is used on the selected robust features for disease prediction. The overview of the method workflow is shown in Fig. \ref{Fig2}.

\subsection{LV segmentation}
We use a U-net shaped architecture (LV-SegNet) for LV segmentation. The architecture of the segmentation model is adapted from \cite{CIBM}. The network takes an echo $\mathcal{I}$ as an input and predict the binary segmentation mask $\mathcal{M}$ as shown in Fig. \ref{Fig2}. 

\subsection{Feature extraction}
The output layer of the encoder (bottleneck) of the LV-SegNet has 32 kernels. Each of the kernels represents a feature activation map corresponding to $\mathcal{I}$. The latent space in the encoder of the LV-SegNet are enriched with geometric features relating to shape, position, texture and topology of the corresponding LV. We use these latent features for feature selection step.

\subsection{Feature selection}
\label{Ft_reduction}
We further reduce the number of latent features by selecting the ones that mostly correlates with the disease diagnosis. Here, we show two independent feature selection technique to reduce the redundant features.

\subsubsection{Version 1: Feature selection using GradCAM kernel ranking (FSR)}

\begin{figure}[h]
	\centering
	\includegraphics[width=3.3in]{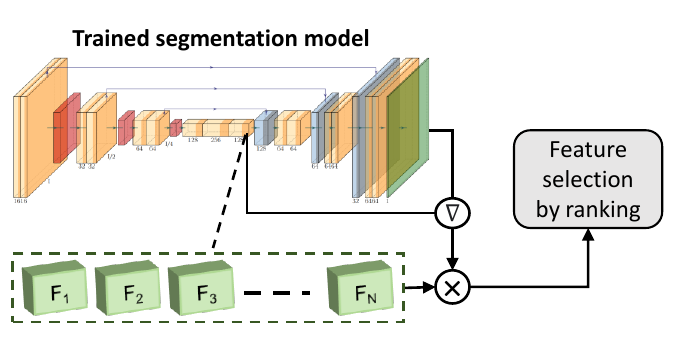}
	\caption{Workflow of the feature selection using GradCAM. Bottleneck kernel weights are obtained through back-propagation. Then the Kernel features are selected based on their weighted ranks and frequency. Finally the reduced feature kernels are flattened and trained with binary learning classifier.} 
	\label{Fig3}
\end{figure}

In our first attempt, we use a feature selection technique based on the GradCAM \cite{gradcam} ranking. Using ground truth and the probabilistic prediction of the LV, we back propagated the loss of LV-SegNet to find out the weights of each of the selected kernels from the previous step. Ideally, the kernel with higher weight has more significance in the downstream segmentation task. The feature reduction workflow is shown in Fig. \ref{Fig3}.

Let, $F^l$, $y^c$, $N$ represents the kernels of the bottleneck, class $c$ related probabilistic output and the number of kernels in the layer of interest, respectively. We can find the weights of the kernels through back propagation using the following equation,

\begin{equation}
    \alpha_{c}^{l} = \frac{1}{N} \sum_{i} \sum_{j} \sum_{k} \frac{\delta y^{c}}{\delta F_{i,j,k}^{l}}
\end{equation}
where $\alpha_{c}^{l}$ represents the kernel weight of $F^l$ corresponding to $y^c$. Finally the GradCAM can be visualized by weighted combination of the feature maps followed by a \textit{ReLU} operation,

\begin{equation}
    L_{c} = \textit{ReLU} \begin{pmatrix}{ 
    \sum_{l} \alpha_{c}^{l} F^{l} }
    \end{pmatrix}
\end{equation}
where, $L^c$ represents the GradCAM of layer $L$ for class $c$. Fig \ref{Fig4}.a)-b) shows sample frame of echo dataset and their corresponding GradCAM visualization of the latent features.

Each of the kernels has weight associated with the LV segmentation. We rank them based on their weights and pick only the $5$ highest weighted kernels for each individual training dataset. Let $S$ be the set of all such kernels. The rank of the kernels can be different for each individual cases due to different activation. We select the $3$ most frequent kernels from the set $S$. Fig \ref{Fig4}.c) shows the accumulated GradCAMs for the 3 most frequent highest weighted kernels. Although,  the latent features are reduced by $90\%$ with this strategy, Fig \ref{Fig4}.c) shows few artifacts are enhanced as well which can be detrimental to train binary classifier for differential diagnosis. 

\subsubsection{Version 2: Feature selection using LASSO algorithm (FSL)}
\begin{figure}[h]
	\centering
	\includegraphics[width=3.3in]{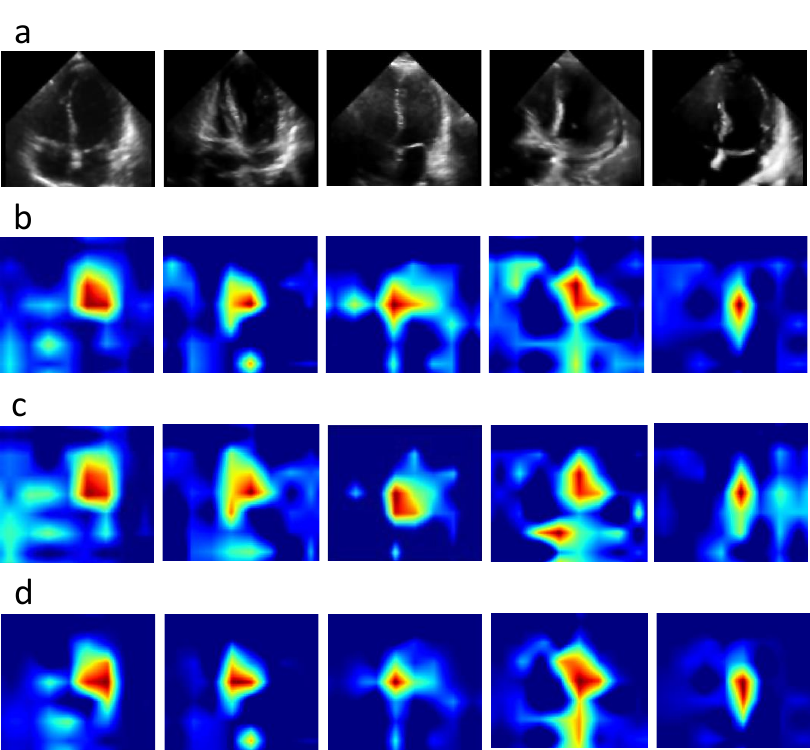}
	\caption{\textbf{a)} Randomly sampled frames of the echo dataset. \textbf{b)} Accumulated GradCAM visualization of $32$ feature kernels of the corresponding echos of a) extracted from the bottleneck. \textbf{c)} Accumulated GradCAM visualization of $3$ feature kernels after selecting feature kernels using weighted ranking. \textbf{d)} Accumulated GradCAM visualization of the feature kernels after selecting features using LASSO. Note that, even with an approximately $90\%$ reduction of the features for c)-d), the highlighted regions are almost identical with b).} 
	\label{Fig4}
\end{figure}

As our ultimate goal is to narrow down the scope of analysis to disease-related robust features, in our second attempt, we used LASSO \cite{lasso1, lasso2} algorithm for feature selection. LASSO as the sparse estimator, is useful for the purpose of feature reduction. The updated weights for each feature can be obtained by minimizing the following objective,

\begin{equation}
    w^{*} = \argmin_{w} { \frac{1}{2n_{\text{samples}}} ||X w - y||_2 ^ 2 + \alpha ||w||_1}
\end{equation}
where ground truths label of the training set denoted by $y$, flattened latent feature vector from bottleneck are denoted by $X$, $w$ denotes the coefficient of regression and $\alpha$ is the penalty coefficient. Fig \ref{Fig4}.d) shows the accumulated GradCAMs of the reduced feature kernels using LASSO.

\subsection{Disease prediction} 
For the binary classification of the selected high-dimensional feature, we investigate 3 independent classifier algorithms:  (1) Support Vector Machine classifier (SVMC), (2) Multi-Layer-Perceptron (MLP), (3) Random forest classifier (RFC).

\subsubsection{SVMC}
We use an SVM classifier with ensemble learning for disease prediction. SVMs are popular and widely used for separating high-dimensional feature space, since it utilizes a subset of training points, referred to as support vectors, in the decision function \cite{cortes1995support}. We used Radial Basis Function (RBF) as SVM kernel while empirically tuning the hyper-parameters. The careful choice of the hyperparameters is pivotal to achieving a balanced hyperplane and avoiding overfitting. Next, to accomplish two key objectives — enhancing prediction accuracy and minimizing spread in predictions — we employed ensemble learning by stacking. Stacking involves training a meta-estimator on the predictions of individual base estimators; it aims to combine the strengths of diverse models and generate a more robust final prediction.

\subsubsection{MLP}
The MLP consists of $3$ fully connected layers having $64$, $256$, and $512$ neurons, respectively. Each fully connected layer followed by a dropout layer to mitigate over-fitting. The output layer, with two neurons for the binary disease class prediction, employs soft-max activation. Binary cross-entropy serves as the loss function.

\subsubsection{RFC}
The Random Forest meta estimator fits a number of decision tree classifiers on various sub-samples of the selected features \cite{random_forest}. In our experiment, we varied the number of estimators and found the optimum estimator size is $100$ and optimum depth of tree size is $11$.

\section{Experiments}
We did 4-fold cross validation for the entire dataset. The proposed segmentation method was implemented using Tensorflow. The network was optimized with the Adam optimizer ($\beta_1=0.9$, $\beta_2=0.999$). The learning rate was initialized as $1e^{-3}$ with an  exponential decay rate of $0.05$ after each epoch. The dataset was normalized to a zero mean and a unit standard variance. To reduce overfitting for the intensity variation of echo, we augmented each echo by adding a number uniformly sampled  in between $-0.1$ and $0.1$ to each image voxel (note that the intensity of each voxel was normalized) for randomly augmenting the video intensities.

We used python scikit-learn library to implement SVMC and RFC. The parameters for SVMC were selected as, (1) Base estimator: $RBF$, (2) meta-estimator: $Logistic Regression$, (3) Learning rate = $1e^{-4}$. The MLP was implemented using Tensorflow with Adam optimizer and with a learning rate $1e^{-3}$.

\section{Results}

\begin{table}[ht!]
    \footnotesize
    \caption{Quantitative results for the differential diagnosis of TTS using echocardiogram video dataset. }
    \label{table1}
    \resizebox{\columnwidth}{!}{
    \begin{tabular}{lcccc}
        \toprule
		\textbf{Method} & \textbf{Sensitivity} & \textbf{Specificity} & \textbf{F1-score} & \textbf{Accuracy} \\
  		\midrule
        \midrule
        DCNN (2D [SCI]) \cite{zaman} & $0.67$ & $0.78$ & $0.69$ & $0.73$ \\
        DCNN (2D [MCI]) \cite{zaman} & $0.73$ & $0.77$ & $0.73$ & $0.75$ \\
        RNN \cite{zaman} & $0.71$ & $0.79$ & $0.72$ & $0.75$ \\
        DCNN (2D+t) \cite{zaman} & $0.79$ & $0.80$ & $0.78$ & $0.80$ \\
        \midrule
        LV-SegNet + SVMC & $0.76$ & $0.84$ & $0.80$ & $0.80$ \\
        LV-SegNet + MLP & $\textbf{0.81}$ & $0.79$ & $0.81$ & $0.80$ \\
        LV-SegNet + RFC & $0.67$ & $0.75$ & $0.71$ & $0.71$ \\
        \midrule
        LV-SegNet + FSR + SVMC & $0.76$ & $0.72$ & $0.75$ & $0.74$ \\
        LV-SegNet + FSR + MLP & $0.75$ & $0.66$ & $0.73$ & $0.71$ \\
        LV-SegNet + FSR + RFC & $0.58$ & $0.75$ & $0.57$ & $0.67$ \\
        \midrule
        LV-SegNet + FSL + SVMC & $0.73$ & $0.82$ & $0.76$ & $0.78$ \\
        LV-SegNet + FSL + MLP & $0.73$ & $\textbf{0.87}$ & $0.80$ & $0.81$ \\
        \textbf{LV-SegNet + FSL + RFC} & $0.78$ & $0.85$ & $\textbf{0.82}$ & $\textbf{0.82}$ \\
        \bottomrule
    \end{tabular}}
\end{table}

We used 4 standard metrics to evaluate disease classification accuracy: (1) Sensitivity, (2) Specificity, (3) F1-score and (4) Accuracy.
Table-\ref{table1} shows the quantitative results with the comparison methods. Results of the DCNN (2D [SCI]), DCNN (2D [MCI]), DCNN (2D+t) and RNN models are obtained from \cite{zaman}. These models directly use echo frame/video for disease prediction. On the other hand, second block of Table-\ref{table1} shows the performance of the $3$ binary classifiers that use the latent features from the LV-SegNet bottleneck. Note that, both SVMC and MLP classifiers using LV-SegNet features have similar performance with the previous SOTA model DCNN(2d+t). Third block of the Table-\ref{table1} shows the performance of classifiers after feature selection using GradCAM kernel ranking. As mentioned before, due to the feature artifacts, these models did not perform well. Fourth block in Table-\ref{table1} shows the performance of classifiers after feature selection using LASSO algorithm. All $3$ classifiers performed well, where RFC outperformed the previous SOTA model with an $82\%$ F1-score and accuracy.

\section{Discussion}
Table-\ref{table1} shows that using the latent space of a segmentation network can achieve better classification accuracy than a direct DCNN classifier. The echocardiogram video dataset is inherently a noisy dataset with lots of speckle noises and artifacts. Hence, the advantage of using latent feature space of a segmentation model is that the features are relatively unbiased to noises and artifacts. Moreover the pathological evidences suggest correlation of region specific features to the TTS. The segmentation network enforces feature learning related to the geometric shape and the motion of LV which plays a significant role in the differential diagnosis. 

The first attempt at reducing redundant features with FSR was unsuccessful. But, it shows that even with $90\%$ reduction of features, we can obtain a reasonable accuracy better than the physicians \cite{zaman}. In the second attempt, the feature reduction technique with LASSO algorithm facilitated robust feature selection that are related to the disease prediction. The binary classifiers with only $10\%$ of the latent features not only improved the overall classification accuracy, but also dramatically reduced the computational complexity.

\section{Conclusion}
Robust feature selection is a crucial step for disease classification and model generalization. Due to the black box nature of the deep-learning models, DCNN classifiers are generally uninterpretable and biased towards noises and artifacts. On the other hand, a segmentation network enforce learning geometric features from the object of interest only, often times specific to disease pathology. Moreover, our feature selection technique shows potential in removing redundant features, thus reducing computational complexity and promoting network interpretability while improving diagnosis accuracy.

\section*{Acknowledgments}
This research was supported in part by the NIH NIBIB Grants R01-EB004640.

\bibliographystyle{IEEEtran}
\bibliography{ISBI2024paper2}

\end{document}